\documentclass[3p,times,procedia]{elsarticle}
\usepackage{nupha_ecrc}

\volume{00}

\firstpage{1}

\journalname{Nuclear Physics A}

\runauth{}

\jid{nupha}

\jnltitlelogo{Nuclear Physics A}

\usepackage{amssymb}

\usepackage[figuresright]{rotating}

\begin{document}

\begin{frontmatter}



\dochead{XXVIth International Conference on Ultrarelativistic Nucleus-Nucleus Collisions\\ (Quark Matter 2017)}

\title{Jet modifications in event-by-event
hydrodynamically evolving media}


\author{Jacquelyn Noronha-Hostler}

\address{Department of Physics, University of Houston, Houston, TX 77204, USA}

\begin{abstract}
The nearly perfect fluid-like nature of the Quark Gluon Plasma may be understood through two key experimental signatures: collective flow and jet suppression.  Event-by-event relativistic viscous hydrodynamics (with an extremely small shear viscosity to entropy density ratio) has been very successful at describing collective flow observables for the last 7 years.  More recently, the effects of event-by-event fluctuations have been studied in the context of high $p_T$ particles that lose energy as they pass through the dense Quark Gluon Plasma liquid. In this summary of the corresponding plenary talk at Quark Matter 2017, the recent developments on the effects of event-by-event fluctuations on jet suppression are summarized.

\end{abstract}


\end{frontmatter}


\section{Introduction}
\label{sec:intro}

The triangular flow study made in relativistic heavy-ion collision in 2010 by \cite{Alver:2010gr}  ignited the event-by-event fluctuating initial conditions ``revolution" within relativistic viscous hydrodynamics \cite{Takahashi:2009na,Schenke:2010rr}.  Since then significant advances have been made in understanding the nature of collective flow, the connection between the initial state and final flow harmonics \cite{Teaney:2010vd,Gardim:2011xv,Gardim:2014tya}, the influence of the Equation of State \cite{Pratt:2015zsa}, and the temperature dependence of various shear and bulk viscosity profiles \cite{Noronha-Hostler:2013gga,Noronha-Hostler:2014dqa,Noronha-Hostler:2015coa,Niemi:2015qia}. It has been shown that the eccentricity $\varepsilon_n$ of the initial state is closely related to the final flow harmonics measured experimentally for elliptical and triangular flow. Thus, if the initial  $\varepsilon_2$ is too small, one would expect difficulties in reproducing large enough $v_2$ or vice versa. 

Theory vs. experimental data comparisons are now so fine-tuned that differences in energies on the order of a few percent points can be predicted \cite{Noronha-Hostler:2015uye,Niemi:2015voa} and later confirmed \cite{Adam:2016izf}.  In the last couple of years, new observables have been created to explore the effects of event-by-event fluctuations such as event-shape engineering  \cite{Adam:2015eta} and Soft-Hard Event Engineering (SHEE) to name a few. Finally, the very nature of the definition of collective flow has come into question due to tantalizing signatures of the Quark Gluon Plasma in small systems \cite{Schenke:2017bog}, so multi-particle cumulants have been used extensively to quantify the strength of fluctuations \cite{Giacalone:2016eyu,Giacalone:2017uqx}. 

Within the initial stages after a heavy-ion collision, pQCD predicts hard scattering processes that produce relativistic jets.  When these relativistic jets pass through the Quark Gluon Plasma they are bumped around by the fluctuations in the medium and lose energy \cite{Gyulassy:2003mc,Majumder:2010qh}. In order to quantify the amount of energy loss transversing the medium the nuclear modification factor, $R_{AA}=\frac{dN_{AA}/dydp_T d\phi}{N_{\rm coll}\,dN_{pp}/dydp_T}$, is measured where the ratio of large systems $AA$ to small systems $pp$ is calculated and one expects values less than 1 when jets are suppressed.   Reconstructing relativistic jets is quite complicated both on the theory and experimental sides but significant advances have been made in Monte Carlo Event Generators \cite{Armesto:2009fj,Young:2011ug,Pang:2012he,Zapp:2013vla,Casalderrey-Solana:2014bpa,Kordell:2016njg} as well as jet reconstruction \cite{jetreview} in recent years.  However, at this point in time, no MC Event Generators are coupled to event-by-event relativistic hydrodynamic codes.  Thus, the rest of this proceedings will focus on high $p_T$ particles as a proxy for jets where various energy loss models are used with a discussion at the end on how MC Event Generators could eventually be coupled to event-by-event relativistic hydrodynamics.  

Below, a discussion follows on how energy loss models have now been combined with event-by-event relativistic hydrodynamic backgrounds to solve the 10 year old $R_{AA}\otimes v_2$ puzzle. Then, theory predictions of high $p_T$ flow harmonics that were confirmed experimentally at LHC run2 are analyzed. Using the knowledge about event-by-event fluctuations Making allows one to think up new experimental observables involving, e.g., Soft-Hard Event Engineering. Finally, I will comment on the possible further implications of initial condition fluctuations on jet observables and the combination of event-by-event hydrodynamics with MC event generators.

\section{$R_{AA}\otimes v_2$ puzzle}
\label{sec:raav2}

While a clear suppression has been measured across all energies from RHIC to LHC, $R_{AA}$ is a relatively robust observable that does not provide a significant amount of distinguishing power between theoretical models.  Thus, Refs.\ \cite{Wang:2000fq,Gyulassy:2000gk} proposed measuring the azimuthal asymmetry of high $p_T$ particles with the understanding that the path length dependence due to an initial anisotropy would lead to a non-zero elliptical flow, $v_2$, at high $p_T$. Indeed, $v_2$ at high $p_T$ was measured \cite{Adler:2005rg} though for $\sim10$ years theoretical calculations consistently under-predicted experimental data \cite{Xu:2014ica} (for a historical overview see \cite{JinfengLiao}).  

\begin{figure*}[ht]
\centering
\includegraphics[width=\textwidth]{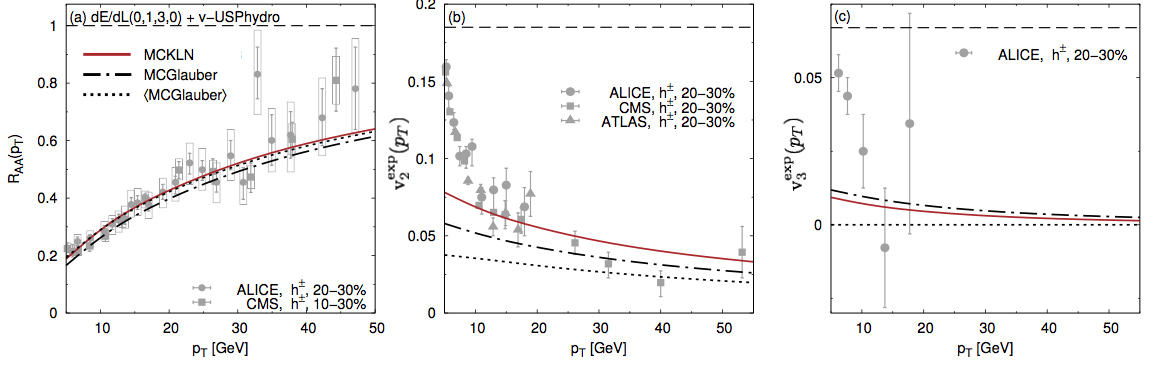}
\caption{(Color online) 
Comparisons of theory calculations (using $dE/dL\sim L$) with various types of initial conditions to data from $20-30\%$ centrality at $\sqrt{s}=2.76$ TeV Pb+Pb collisions at the LHC \cite{Abelev:2012hxa,CMS:2012aa,Abelev:2012di,Chatrchyan:2012xq,Aad2012330}. The experimental observables are (a) the nuclear modification factor $R_{AA}$, (b) elliptical flow $v_2$, and (c) triangular flow $v_3$. MCKLN initial conditions are shown in the solid red lines, MCGlauber is black dashed-dotted, and averaged MCGlauber in black dotted neglects initial state fluctuations. }
\label{fig:RAAv2v3}
\end{figure*}

In \cite{Noronha-Hostler:2016eow} the first calculations combining event-by-event viscous hydrodynamics with an energy loss model (v-USPhydro+BBMG \cite{Noronha-Hostler:2013gga,Noronha-Hostler:2014dqa,Noronha-Hostler:2015coa,Betz:2011tu,Betz:2012qq}) were made where $R_{AA}$, $v_2(p_T)$, and $v_3(p_T)$ gave a good description of experimental data, as shown in Fig.\ \ref{fig:RAAv2v3}. All previous calculations using averaged initial conditions returned $v_3=0$ and only in the case that event-by-event fluctuations are considered can one obtain a non-zero $v_3$ across all $p_T$'s. 

Experimentally, $v_n(p_T)$ is measured using the scalar product \cite{Bilandzic:2010jr} 
\begin{equation}\label{eqn:vncor}
v_n\{2\}(p_T)=\frac{\langle v_n\,v_n^{hard}(p_T)\cos\left[n\left(\psi_n-\psi^{hard}_n(p_T)\right]\right) \rangle}{\sqrt{\left\langle \left(v_n\right)^{2}\right\rangle}}.
\end{equation}
where $v_n$ is the integrated flow harmonic in the soft sector and $v_n^{hard}(p_T) = \frac{\frac{1}{2\pi}\int_0^{2\pi}d\phi\,\cos\left[n\phi-n\psi_n^{hard}(p_T)\right]\,R_{AA}(p_T,\phi)}{R_{AA}(p_T)}$.
Statistics prevents one to measure two high $p_T$ particles and correlate them as is typically done to measure $v_n\{2\}$ and, thus, $v_n\{SP\}(p_T)$ correlates one soft particle and one hard particle. Due to this soft-hard correlation, it is vital to first ensure that the soft flow harmonics match well to experimental data before one attempts to calculate $v_n(p_T>10\,\mathrm{GeV})$. 

Drawing on previous results in the soft sector, one can conclude that the choice in event-by-event fluctuating initial conditions play a vital role in ensuring that theoretical calculations can match experimental data. In our calculations there is roughly a $30\%$ difference between the eccentricities, $\varepsilon_2$, of MCKLN and MCGlauber, which corresponds to the roughly $30\%$ larger $v_2$ shown in Fig.\ \ref{fig:RAAv2v3}. Triangular flow is also shown in Fig.\ \ref{fig:RAAv2v3} where the values are quite small.  One can understand that due to the decorrelation of the event plane angles i.e. the high $p_T$ particle's angle is no longer aligned with the soft event plane angle \cite{Jia:2012ez,Betz:2016ayq}.  For $v_2$ the angles are quite closely aligned though for higher order flow harmonics the decorrelation effect becomes stronger. 

\section{LHC Run 2}
\label{sec:run2}

\begin{figure*}[ht]
\centering
\includegraphics[width=0.4\textwidth]{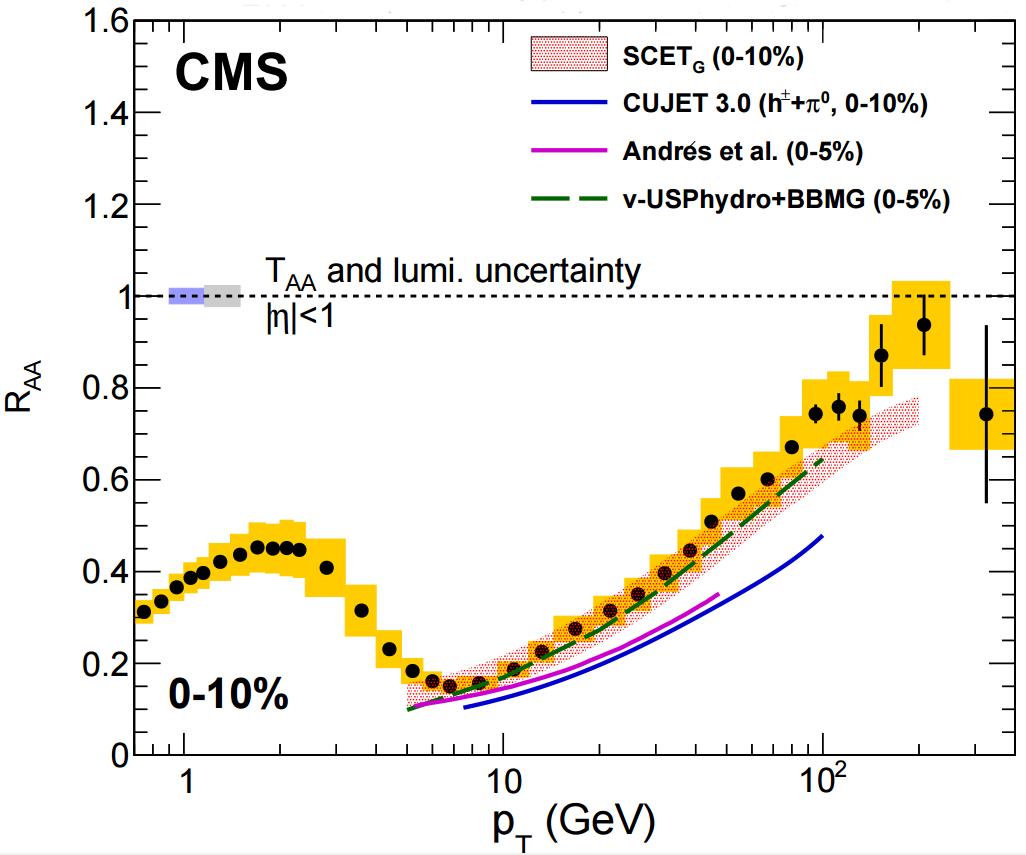}  
\caption{(Color online) $R_{AA}$ predictions from v-USPhydro+BBMG \cite{Betz:2016ayq}, SCET \cite{Chien:2015vja}, CUJET3.0 \cite{JinfengLiao}, and Andres et al \cite{Andres:2016iys} compared to CMS at LHC $\sqrt{s_{NN}}=5.02$ TeV  from \cite{Khachatryan:2016odn}. }
\label{fig:RAArun2}
\end{figure*}

The larger luminosity at LHC Run 2 provides a unique opportunity to further test the relevance of event-by-event fluctuations on high $p_T$ observables with significantly smaller error bars.  In Fig.\ \ref{fig:RAArun2} the $R_{AA}$ at LHC run 2 is shown compared to various theoretical models \footnote{Other preliminary experimental $R_{AA}$ results for run 2 can be found for ATLAS \cite{ATLAS-CONF-2017-012.} and ALICE \cite{Huhn:2017uus}}. v-USPHYDR+BBMG \cite{Betz:2016ayq}   and SCET \cite{Chien:2015vja} (not shown: a revised version of CUJET3.0 \cite{JinfengLiao}, a scale dependent $\hat{q}$ model \cite{Bianchi:2017wpt}, and a radiative+collisional energy loss model \cite{Djordjevic:2016vfo}) are able to match experimental data well. 

Alternatively, the flow harmonics have been found to be sensitive to the choice of initial conditions and event-by-event fluctuations \cite{Noronha-Hostler:2016eow}. While the dominant factor appears to be the choice of initial conditions themselves (this stresses the importance of first fitting both $v_2\{2\}$ and $v_3\{2\}$ in the soft sector before making predictions at high $p_T$), event-by-event fluctuations and experimental effects such as centrality rebinning and multiplicity weighing have also been taken into account in theoretical calculations \cite{Betz:2016ayq}. In the era of precision data from LHC run 2 then it is necessary to include these effects in theoretical calculations in order to extract information about the energy loss properties.  

\begin{figure*}[ht]
\centering
\includegraphics[width=1\textwidth]{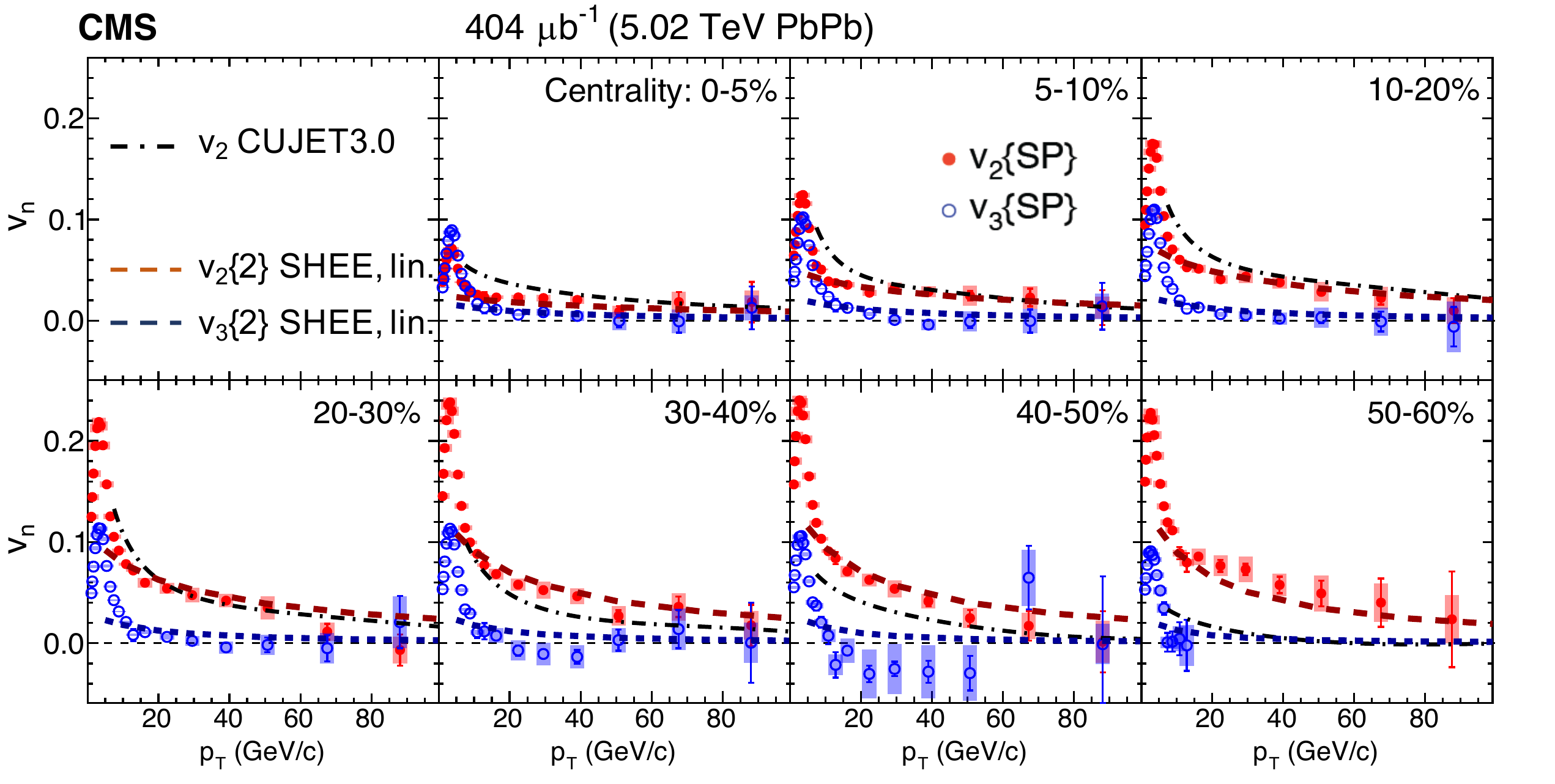}  
\caption{(Color online) $v_2\{SP\}(p_T)$ and $v_3\{SP\}(p_T)$ data from CMS at LHC $\sqrt{s_{NN}}=5.02$ TeV  from \cite{Khachatryan:2016odn} compared to $v_2\{SP\}(p_T)$ and $v_3\{SP\}(p_T)$ calculations from v-USPhydro+BBMG (SHEE) \cite{Betz:2016ayq} and $v_2^{hard}(p_T)$ calculations from CUJET3.0 \cite{JinfengLiao}. }
\label{fig:vnrun2}
\end{figure*}

In Fig.\ \ref{fig:vnrun2} two theory calculations are compared to the latest $v_2\{SP\}(p_T)$ and $v_3\{SP\}(p_T)$ data from CMS at LHC $\sqrt{s_{NN}}=5.02$ TeV.  v-USPhydro+BBMG includes event-by-event fluctuations so within this framework it is possible to calculate the full scalar product with centrality rebinning and multiplicity weighing \cite{Betz:2016ayq}. CUJET3.0 uses hydrodynamical backgrounds from averaged initial conditions and, therefore, can only calculate $v_2^{hard}(p_T)$. For $dE/dL\sim L$, v-USPhydro+BBMG shows a very good agreement to experimental data (while $dE/dL\sim L^2$ overpredicts data, not shown).  However, CUJET3.0 slightly overpredicts $v_2$ for central collisions and largely underpredicts it for peripheral collisions.  Currently, the CUJET3.0 group is working on incorporating event-by-event fluctuations and results should be out shortly \cite{JinfengLiao}. Not shown in the plots were $v_2^{hard}(p_T)$ calculations from \cite{Bianchi:2017wpt} for $0-40\%$ centrality with averaged initial conditions.  

At this conference, data for $v_2\{SP\}(p_T)$-$v_4\{SP\}(p_T)$ were also shown by ALICE \cite{Bertens:2017krr} up to $p_T=40$ GeV while results for $v_2\{SP\}(p_T)$-$v_7\{SP\}(p_T)$ up to $p_T=25$ GeV were shown by ATLAS \cite{ATLAS-CONF-2016-105}. It does appear that higher order flow harmonics $n\geq 4$ are essentially consistent with zero above $p_T\geq 10$ GeV, which can be explained again by the decorrelation of the event plane angles \cite{Jia:2012ez}. 

\begin{figure*}[ht]
\centering
\includegraphics[width=0.4\textwidth]{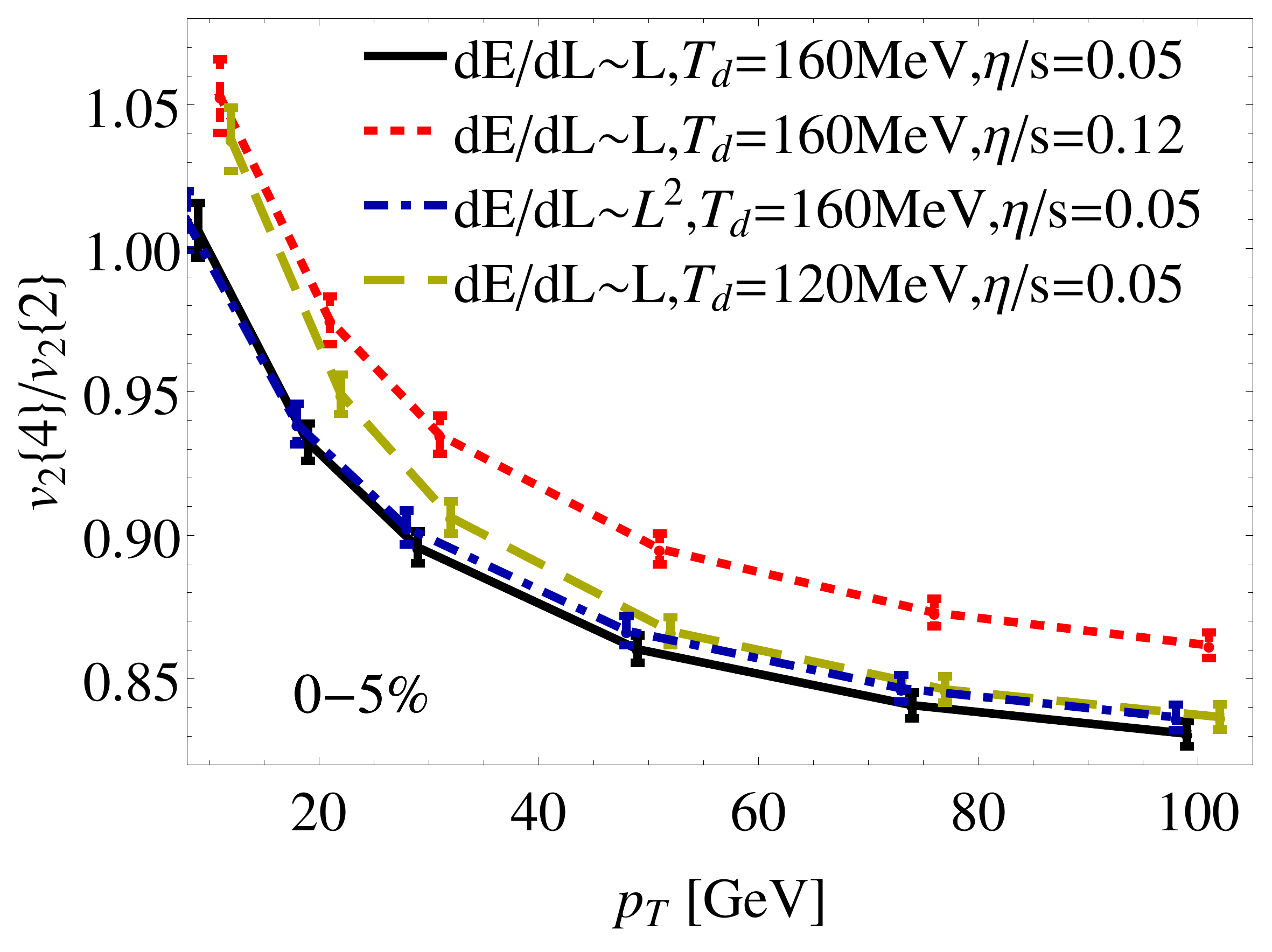}  
\caption{(Color online) Ratio of $\frac{v_n\{4\}(p_T) }{v_n\{2\}(p_T) }$ with varying $\eta/s$, $T_d$, and $dE/dL$ from \cite{Betz:2016ayq} at LHC $\sqrt{s_{NN}}=5.02$ TeV.  }
\label{fig:cumrun2}
\end{figure*}

At LHC run2 multiplicity cumulants at high $p_T$ have now been measured where one high $p_T$ is correlated with an odd number of soft particles i.e. $v_2\{4\}$ involves one hard particle and 3 soft particles. v-USPhydro+BBMG \cite{Betz:2016ayq} made predictions for LHC run2, which agreed well with measurements by CMS \cite{QWang}.  One surprising consequence of this measurement is that the ratio of $v_2\{4\}/v_2\{SP\}(p_T)\sim 1$ at $p_T\geq 10$ GeV as shown in Fig. \ref{fig:cumrun2}.  This arises because of the soft-hard correlation of the multi-particle scalar product such that
\begin{equation}\label{eqn:v4}
\frac{v_n\{4\}(p_T) }{v_n\{2\}(p_T) }=\frac{v_n\{4\} }{v_n\{2\} } \left[1+\left(\frac{v_n\{2\}}{v_n\{4\}}\right)^4\underbrace{\left(\frac{\langle  v_n^4\rangle}{\langle  v_n^2\rangle^2}- \frac{\langle v_n^2  V_n V_n^*(p_T)\rangle}{\langle v_n^2\rangle\langle  V_n V_n^*(p_T)\rangle}\right)}_{\textit{soft-hard fluctuations}}  \right],   \label{eqn:corterm}
\end{equation}
where $V_n$ is the flow vector that includes both its magnitude and event plane angle. 
Eq.\ (\ref{eqn:v4}) can be derived directly from the equations for the scalar product \cite{Betz:2016ayq}. Essentially, if the fluctuations in the soft and hard sector are the same then $\frac{v_n\{4\}(p_T) }{v_n\{2\}(p_T) }=\frac{v_n\{4\} }{v_n\{2\} } $.  When $\frac{v_n\{4\}(p_T) }{v_n\{2\}(p_T) }>\frac{v_n\{4\} }{v_n\{2\} } $, it is an indication that fluctuations in the hard sector begin to dominate, which occurs right around $p_T\sim 10$ GeV according to CMS data \cite{QWang} and theory \cite{Betz:2016ayq}. These results highlight the importance of both event-by-event fluctuations in the soft sector as well as intrinsic jet fluctuations in the hard sector in order to describe the multiparticle cumulants across $p_T$.

\section{Soft-Hard Event Engineering}
\label{sec:SHEE}

\begin{figure*}[ht]
\centering
\includegraphics[width=\textwidth]{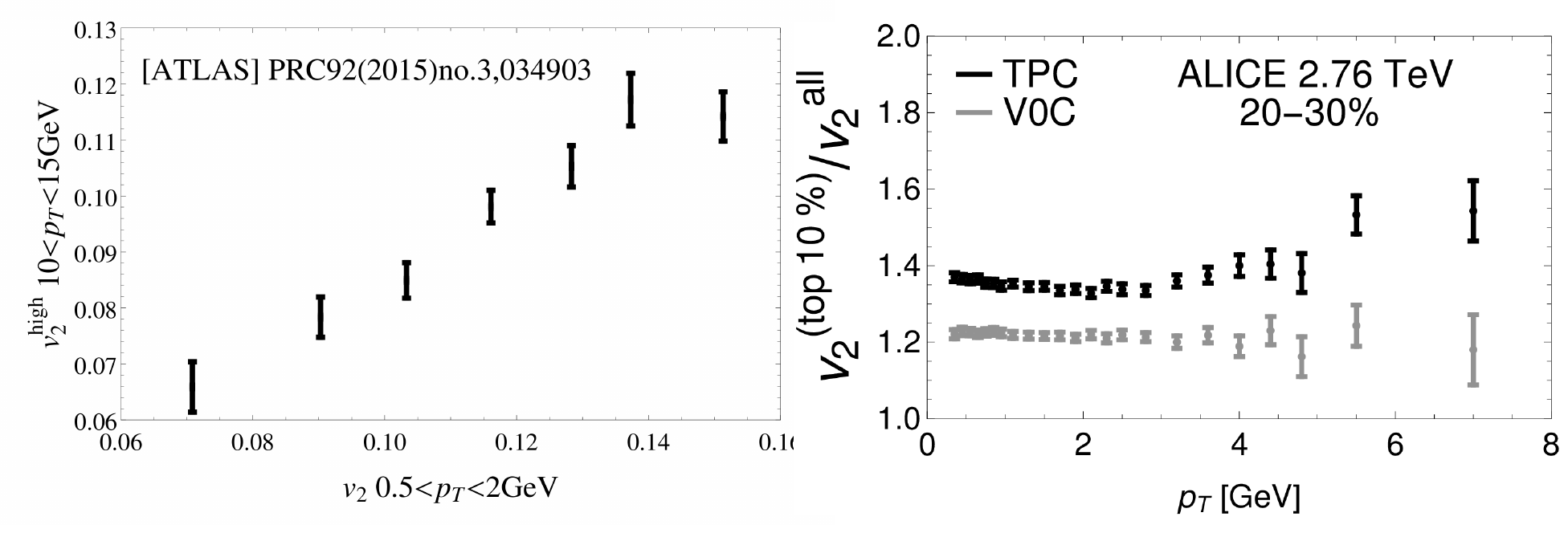}  
\caption{(Color online) Compilation of Soft-Hard Event Engineering measurements from ATLAS \cite{Aad:2015lwa} and ALICE \cite{Adam:2015eta} at LHC $\sqrt{s_{NN}}=2.76$ TeV.}
\label{fig:expSHEE}
\end{figure*}
A major consequence of event-by-event fluctuations is the ability to explore high $p_T$ observables using Soft-Hard Event Engineering (SHEE) i.e calculating high $p_T$ observables while selecting the event shape in soft sector (flow harmonics). Fig.\ \ref{fig:expSHEE} shows a compilation of two different examples of SHEE that have already been measured above $p_T>10$ GeV at LHC $\sqrt{s_{NN}}=2.76$ TeV within the $20-30\%$ centrality class. In \cite{Aad:2015lwa} ATLAS sorted the integrated $v_n\{2\}$ in the soft sector and then measured the corresponding  $v_n\{SP\}$ at high $p_T$ within each bin shown in Fig.\ \ref{fig:expSHEE} (left).  One can see that there is a clear scaling with the underlying event shape event up to $p_T\sim 15$ GeV.  This is unambiguous evidence that event-by-event fluctuations affect high $p_T$ flow harmonics (if high $p_T$ flow harmonics were unaffected one would see a flat line here).  In Fig.\ \ref{fig:expSHEE} (right) the top $10\%$ largest $v_2\{2\}$ are selected integrated over all $p_T$ and then the corresponding $v_n^{10\%}\{SP\}(p_T)/v_n^{all}\{SP\}(p_T)$ is plotted \cite{Adam:2015eta}.  The fact that the enhancement is seen universally across all $p_T$ also clearly indicates that the flow across all $p_T$ originates from the same source (initial eccentricities). 

\begin{figure*}[ht]
\centering
\begin{tabular}{c c c}
\includegraphics[width=0.33\textwidth]{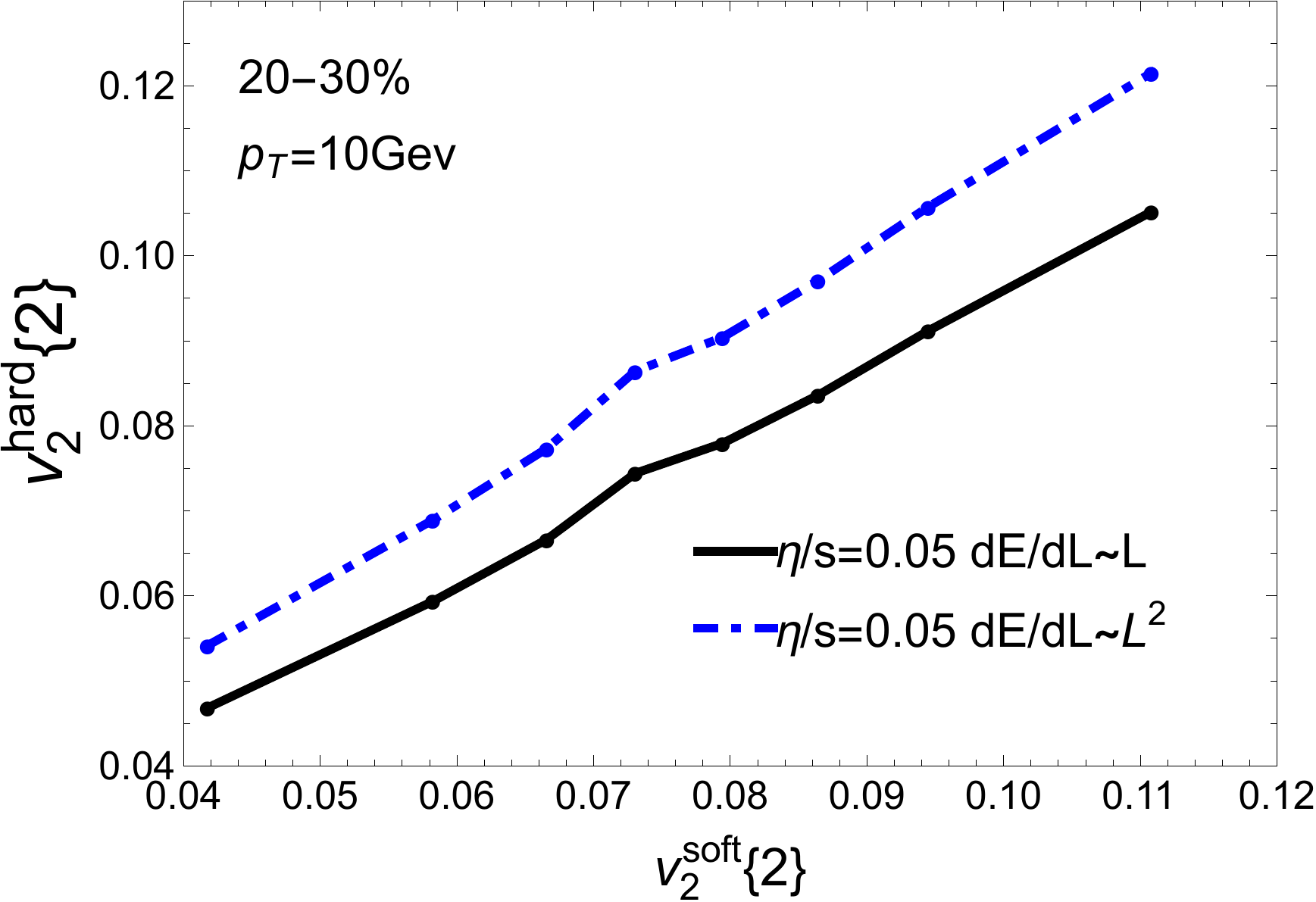}  & \includegraphics[width=0.33\textwidth]{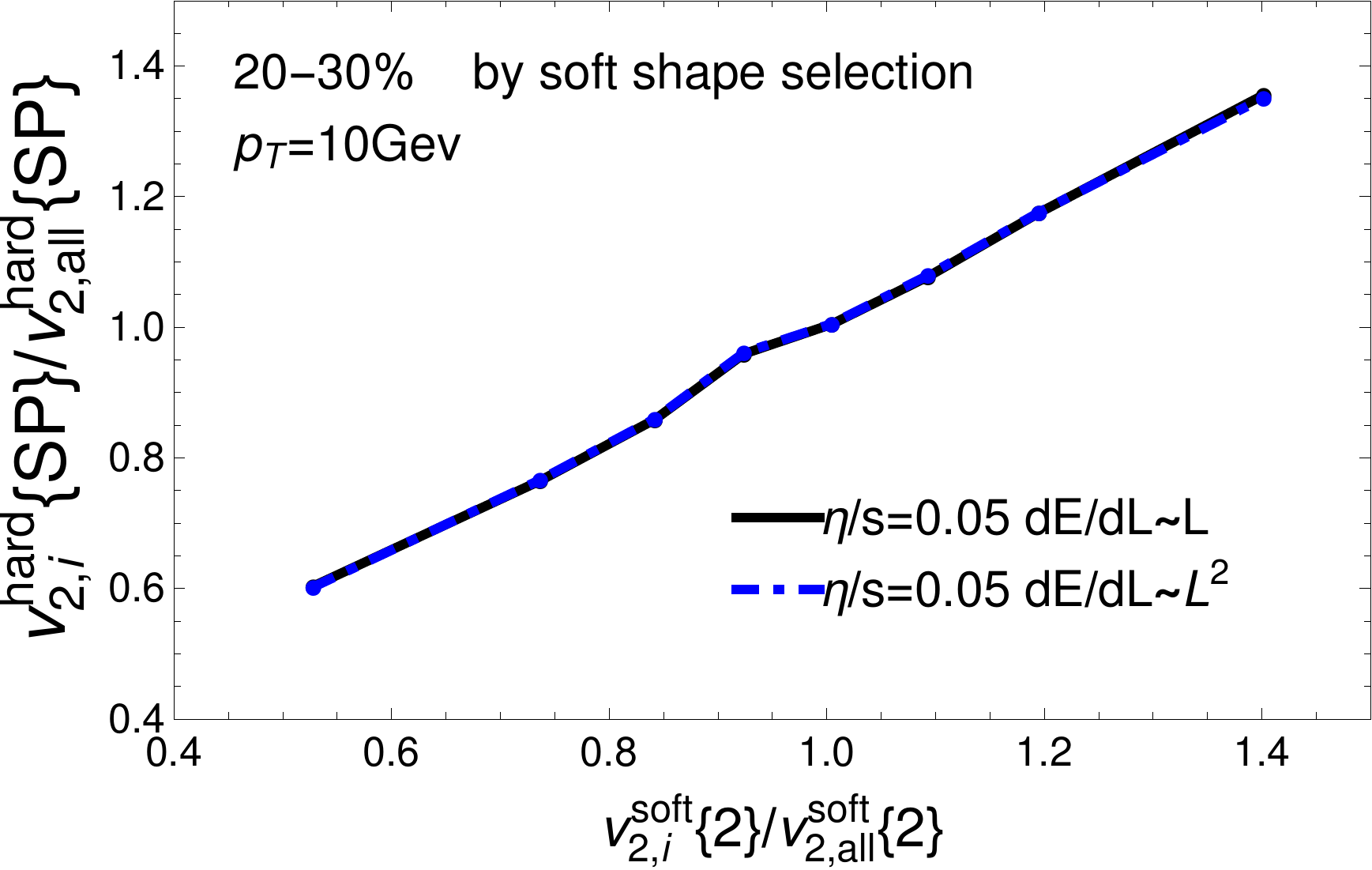} & \includegraphics[width=0.33\textwidth]{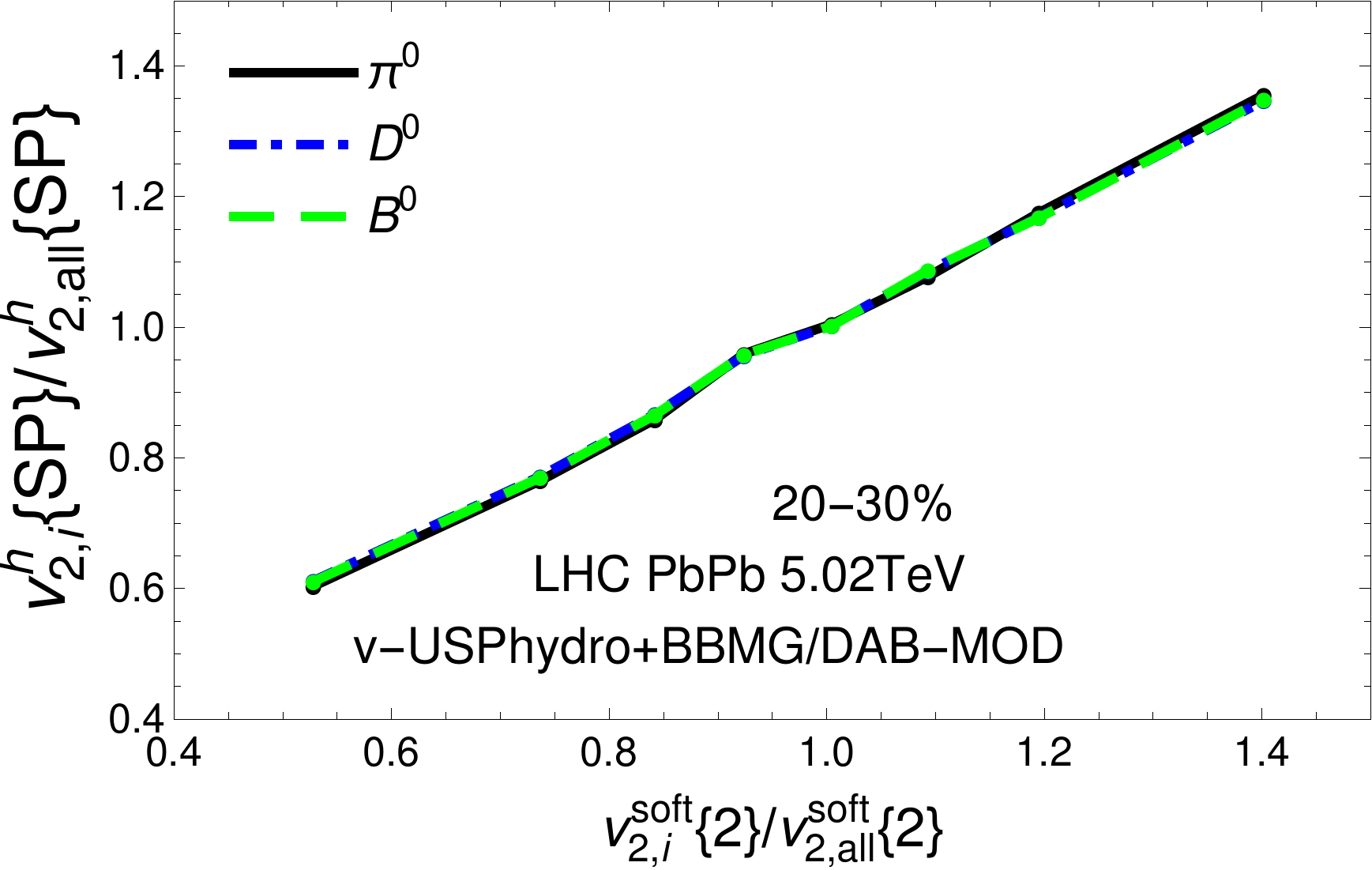}
\end{tabular}
\caption{(Color online) Predictions for SHEE of the magnitude of $v_2$ at $20-30\%$ LHC $\sqrt{s_{NN}}=5.02$ TeV computed using two different $dE/dL$'s (left), rescaled by the total $v_2$ for $20-30\%$ for light hadrons (middle), and a comparison of the rescaled SHEE result for different flavors (right). }
\label{fig:universal}
\end{figure*}

The first theoretical calculations of this type are shown in Fig.\ \ref{fig:universal} (left) at LHC $\sqrt{s_{NN}}=5.02$ TeV where the same methodology as ATLAS \cite{Aad:2015lwa} was  used. Events in the $20-30\%$ centrality class were sorted into 8 subevents depending on the soft $v_2\{2\}$ values and then the corresponding $v_2\{SP\}$ at $p_T=10$ GeV was calculated.  One can see that there is a very clear linear scaling of the calculation with the soft sector, which means that events with a large eccentricity produce high $p_T$ events with a large $v_2$ (and vice versa).  Additionally, the slope in Fig.\ \ref{fig:jedmed} (left) is strongly dependent on the energy loss model details (e.g., $dE/dL$).  One could imagine that something like $dE/dL\sim L^3$ would be even larger with a steeper slope.  Thus, this type of measurement could give strong insight into the path length dependence of energy loss seen in the Quark Gluon Plasma.

A natural extension of this would be then to rescale Fig.\ \ref{fig:universal} (left) by $v_2\{2\}$ for all events in $20-30\%$ along the horizontal axis and by $v_2\{SP\}$  for all events in $20-30\%$ along the vertical axis. Doing so one arrives at Fig.\ \ref{fig:universal} (middle, right) where the dependence on the choice of energy loss model and even flavor dependence disappears ($D^0$ and $B^0$ mesons were calculated using v-USPhydro+DAB-MOD \cite{Prado:2016szr}). This universal scaling could be readily verified experimentally using LHC run2 data and it would provide conclusive evidence that flow harmonics, across all species, originate from the same source (the underlying initial state eccentricities).

\section{Outlook}
\label{sec:out}

The first experimental results that connected both soft and hard fluctuations occurred less than two years ago \cite{Aad:2015lwa,Adam:2015eta} and the first theory calculations are only about a year old \cite{Noronha-Hostler:2016eow,Betz:2016ayq}. Therefore, there are many new possibilities to be explored both experimentally and theoretically in this area.  Thus far, we have learned that $R_{AA}$ is insensitive to event-by-event fluctuations arising from the soft sector whereas $v_n(p_T)$ for $p_T>10$ GeV is a combination of event-by-event initial condition fluctuations and intrinsic jet energy loss fluctuations.  However, at this point many opportunities remain to calculate fully reconstructed jets observables on an event-by-event basis to see if there is any underlying effect from the eccentricities of the initial conditions.  For instance, any sort of correlations across $p_T$ could also be extrapolated upward to higher $p_T$ as well \cite{Qian:2017ier}.

\begin{figure*}[ht]
\centering
\begin{tabular}{c c}
\includegraphics[width=0.5\textwidth]{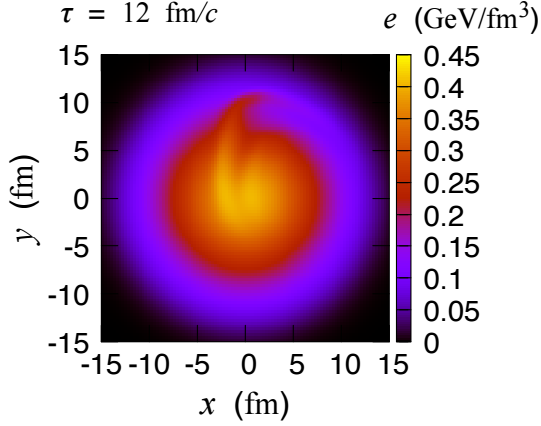}  & \includegraphics[width=0.45\textwidth]{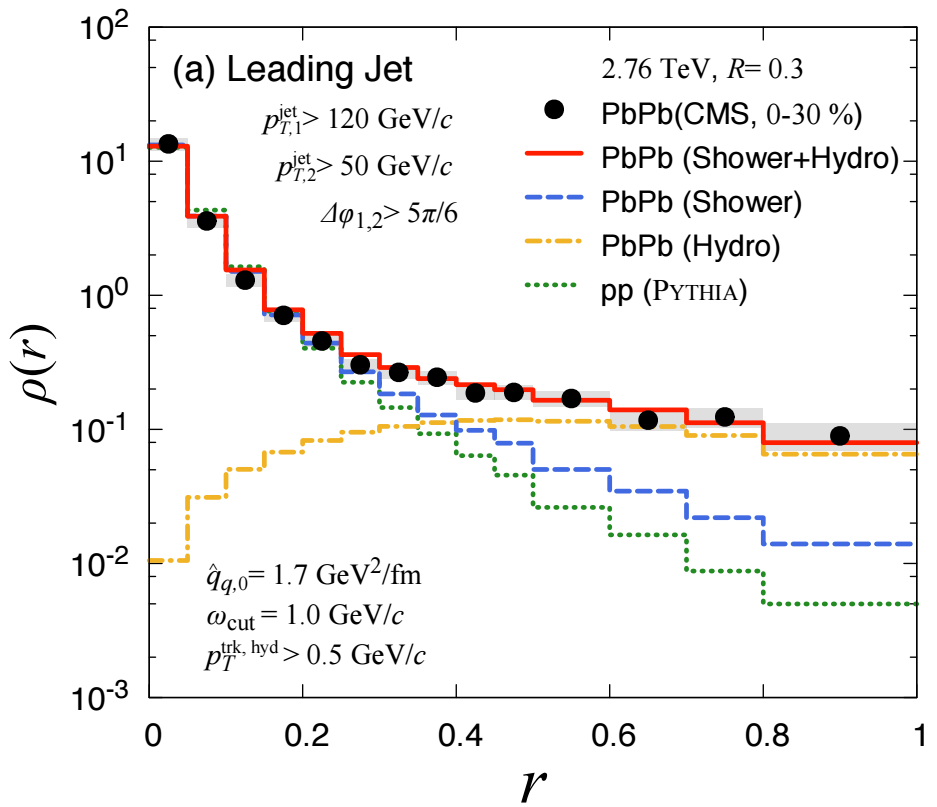}
\end{tabular}
\caption{(Color online) Jet embedded in an optical Glauber event (left) and the effect of the medium on the jet shape function (right). Figure taken with the permission of the authors from \cite{Tachibana:2017syd}.}
\label{fig:jedmed}
\end{figure*}
One possible path forward when it comes to full jet reconstruction on an event-by-event basis involves including a source term in relativistic hydrodynamics such that jets can be embedded directly into the medium \cite{Pang:2012he,Andrade:2014swa,Tachibana:2017syd}.  At Quark Matter 2017 results using an optical Glauber initial condition were shown where jets were embedded into the medium using a source term in 3+1 ideal hydrodynamics.  In Fig.\ \ref{fig:jedmed} one can see a picture of the jet traversing the medium (left) and the effects that the medium has on the jet shape function (right), according to Ref.\ \cite{Tachibana:2017syd}. The jet shape function sees a significant improvement when the effect of the medium is incorporated.  In such an approach, it would be possible to eventually use the same jet finding algorithms in both theory and experiment, which would constitute a significant step forward towards understanding reconstructed jets. There are a number of groups tackling the issue of combining event-by-event hydrodynamics with jets.  A natural framework for this is the NSF funded JETSCAPE collaboration whose aim is to improve computational techniques of Monte Carlo Event Generators in order to make them publicly available and easily usable.  In Europe THOR also has a working group on Hard Probes that could tackle some of these problems. 

While much of this proceedings was focused on light high $p_T$ particles, one would be remiss not to point out the significant progress made in the heavy flavor sector where event-by-event initial condition fluctuations have already been taken into account for a few years. The first event-by-event fluctuating initial conditions were used within an ideal hydrodynamic setup by the SUBATECH group in \cite{Nahrgang:2014vza,Nahrgang:2016wig} where $v_2(p_T)-v_4(p_T)$ were calculated using the event plane method.  In \cite{Cao:2013ita} a large effect on the heavy flavor flow was seen when the initial conditions were varied (using averaged initial conditions for the event plane method). In \cite{Prado:2016szr} the first Soft-Heavy Event Engineering theoretical calculations were made and ALICE has begun incorporating SHEE into the heavy flavor as well where \cite{Barbano} showed  the D meson analog of Fig.\ \ref{fig:expSHEE} (right). Additionally, the use of Bayesian analysis in the heavy flavor sector has resulted in a more constrained understanding of transport coefficients \cite{YingruXU}. For the future, exploring the flavor mass dependence in event shape engineering could be quite interesting \cite{Christiansen:2016uaq}.

Finally, a puzzle still remains in the intermediate $p_T$ region in terms of how to interpolate correctly between the soft and hard regions of physics.  At this point, relativistic  hydrodynamics is either taken as a background for the hard sector or in the soft sector jets are included as a source  term within hydrodynamics.  However, it may be that the only possible solution to the intermediate $p_T$ region involves jets fully incorporated into hydrodynamics include all the possible backreaction effects.  Such an endeavor would be extremely computational intensive because having one jet per event would require statistics on the same order of experiments.  Additionally, all flow calculations would then need to be done with $Q_n$ vectors used in experiments \cite{Bilandzic:2010jr} with a rapidity gap to perform an apples-to-apples comparisons.  However, a fully integrated hydrodynamics+jets code would be well worth the effort to guide the new age of precision measurements at LHC and the upcoming sPHENIX experiment at RHIC.

\section{Acknowledgements}
\label{sec:awk}

JNH would like to thank A.\ Angerami, B.\ Betz, M.\ Gyulassy, W.\ Li, M.\ Luzum, S.\ Mohapatra, M.\ Munhoz, C.\ Nattrass, J.\ Noronha, C.\ Prado, C.\ Ratti, A.\ Suaide, A.\ Timmins, and Q.\ Wang for discussions on this topic as well as all the other speakers who provided figures and contributed content. J.N.H. was supported by the National Science Foundation under grant no. PHY-1513864 and she acknowledges the use of the Maxwell Cluster and the advanced support from the Center of Advanced Computing and Data Systems at the University of Houston to carry out the research presented here.

\bibliographystyle{elsarticle-num}
\bibliography{library}

\begin{thebibliography}{10}
\expandafter\ifx\csname url\endcsname\relax
  \def\url#1{\texttt{#1}}\fi
\expandafter\ifx\csname urlprefix\endcsname\relax\def\urlprefix{URL }\fi
\expandafter\ifx\csname href\endcsname\relax
  \def\href#1#2{#2} \def\path#1{#1}\fi

\bibitem{Alver:2010gr}
B.~Alver, G.~Roland, Phys. Rev. C81 (2010) 054905, [Erratum: Phys.
  Rev.C82,039903(2010)].

\bibitem{Takahashi:2009na}
J.~e.~a. Takahashi, Phys. Rev. Lett. 103 (2009) 242301.
\newblock \href {http://dx.doi.org/10.1103/PhysRevLett.103.242301}
  {\path{doi:10.1103/PhysRevLett.103.242301}}.

\bibitem{Schenke:2010rr}
B.~Schenke, S.~Jeon, C.~Gale, Phys. Rev. Lett. 106 (2011) 042301.
\newblock \href {http://dx.doi.org/10.1103/PhysRevLett.106.042301}
  {\path{doi:10.1103/PhysRevLett.106.042301}}.

\bibitem{Teaney:2010vd}
D.~Teaney, L.~Yan, Phys. Rev. C83 (2011) 064904.
\newblock \href {http://dx.doi.org/10.1103/PhysRevC.83.064904}
  {\path{doi:10.1103/PhysRevC.83.064904}}.

\bibitem{Gardim:2011xv}
F.~G. Gardim, F.~Grassi, M.~Luzum, J.-Y. Ollitrault, Phys. Rev. C85 (2012)
  024908.

\bibitem{Gardim:2014tya}
F.~G. Gardim, J.~Noronha-Hostler, M.~Luzum, F.~Grassi, Phys. Rev. C91~(3)
  (2015) 034902.

\bibitem{Pratt:2015zsa}
S.~Pratt, E.~Sangaline, P.~Sorensen, H.~Wang, Phys. Rev. Lett. 114 (2015)
  202301.

\bibitem{Noronha-Hostler:2013gga}
J.~Noronha-Hostler, G.~S. Denicol, J.~Noronha, R.~P.~G. Andrade, F.~Grassi,
  Phys. Rev. C88~(4) (2013) 044916.

\bibitem{Noronha-Hostler:2014dqa}
J.~Noronha-Hostler, J.~Noronha, F.~Grassi, Phys. Rev. C90~(3) (2014) 034907.

\bibitem{Noronha-Hostler:2015coa}
J.~Noronha-Hostler, J.~Noronha, M.~Gyulassy, Phys. Rev. C93~(2) (2016) 024909.

\bibitem{Niemi:2015qia}
H.~Niemi, K.~J. Eskola, R.~Paatelainen, Phys. Rev. C93~(2) (2016) 024907.

\bibitem{Noronha-Hostler:2015uye}
J.~Noronha-Hostler, M.~Luzum, J.-Y. Ollitrault, Phys. Rev. C93~(3) (2016)
  034912.

\bibitem{Niemi:2015voa}
H.~Niemi, K.~J. Eskola, R.~Paatelainen, K.~Tuominen, Phys. Rev. C93~(1) (2016)
  014912.
\newblock \href {http://dx.doi.org/10.1103/PhysRevC.93.014912}
  {\path{doi:10.1103/PhysRevC.93.014912}}.

\bibitem{Adam:2016izf}
J.~Adam, et~al., Phys. Rev. Lett. 116~(13) (2016) 132302.
\newblock \href {http://arxiv.org/abs/1602.01119} {\path{arXiv:1602.01119}},
  \href {http://dx.doi.org/10.1103/PhysRevLett.116.132302}
  {\path{doi:10.1103/PhysRevLett.116.132302}}.

\bibitem{Adam:2015eta}
J.~Adam, et~al., Phys. Rev. C93~(3) (2016) 034916.
\newblock \href {http://dx.doi.org/10.1103/PhysRevC.93.034916}
  {\path{doi:10.1103/PhysRevC.93.034916}}.

\bibitem{Schenke:2017bog}
B.~Schenke\href {http://arxiv.org/abs/1704.03914} {\path{arXiv:1704.03914}}.

\bibitem{Giacalone:2016eyu}
G.~Giacalone, L.~Yan, J.~Noronha-Hostler, J.-Y. Ollitrault, Phys. Rev. C95~(1)
  (2017) 014913.
\newblock \href {http://dx.doi.org/10.1103/PhysRevC.95.014913}
  {\path{doi:10.1103/PhysRevC.95.014913}}.

\bibitem{Giacalone:2017uqx}
G.~Giacalone, J.~Noronha-Hostler, J.-Y. Ollitrault\href
  {http://arxiv.org/abs/1702.01730} {\path{arXiv:1702.01730}}.

\bibitem{Gyulassy:2003mc}
M.~Gyulassy, I.~Vitev, X.-N. Wang, B.-W. Zhang\href
  {http://arxiv.org/abs/nucl-th/0302077} {\path{arXiv:nucl-th/0302077}}, \href
  {http://dx.doi.org/10.1142/9789812795533 0003}
  {\path{doi:10.1142/9789812795533 0003}}.

\bibitem{Majumder:2010qh}
A.~Majumder, M.~Van~Leeuwen, Prog. Part. Nucl. Phys. 66 (2011) 41--92.
\newblock \href {http://dx.doi.org/10.1016/j.ppnp.2010.09.001}
  {\path{doi:10.1016/j.ppnp.2010.09.001}}.

\bibitem{Armesto:2009fj}
N.~Armesto, L.~Cunqueiro, C.~A. Salgado, Eur. Phys. J. C63 (2009) 679--690.

\bibitem{Young:2011ug}
C.~Young, B.~Schenke, S.~Jeon, C.~Gale, Phys. Rev. C86 (2012) 034905.

\bibitem{Pang:2012he}
L.~Pang, Q.~Wang, X.-N. Wang, Phys. Rev. C86 (2012) 024911.

\bibitem{Zapp:2013vla}
K.~C. Zapp, Eur. Phys. J. C74~(2) (2014) 2762.

\bibitem{Casalderrey-Solana:2014bpa}
J.~Casalderrey-Solana, D.~C. Gulhan, J.~G. Milhano, D.~Pablos, K.~Rajagopal,
  JHEP 10 (2014) 019, [Erratum: JHEP09,175(2015)].

\bibitem{Kordell:2016njg}
M.~Kordell, A.~Majumder\href {http://arxiv.org/abs/1601.02595}
  {\path{arXiv:1601.02595}}.

\bibitem{jetreview}
M.~Connors, C.~Nattrass, R.~Reed, S.~Salur, {To appear shortly}.

\bibitem{Wang:2000fq}
X.-N. Wang, Phys. Rev. C63 (2001) 054902.
\newblock \href {http://dx.doi.org/10.1103/PhysRevC.63.054902}
  {\path{doi:10.1103/PhysRevC.63.054902}}.

\bibitem{Gyulassy:2000gk}
M.~Gyulassy, I.~Vitev, X.~N. Wang, Phys. Rev. Lett. 86 (2001) 2537--2540.
\newblock \href {http://dx.doi.org/10.1103/PhysRevLett.86.2537}
  {\path{doi:10.1103/PhysRevLett.86.2537}}.

\bibitem{Adler:2005rg}
S.~S. Adler, et~al., Phys. Rev. Lett. 96 (2006) 032302.

\bibitem{Xu:2014ica}
J.~Xu, A.~Buzzatti, M.~Gyulassy, JHEP 08 (2014) 063.
\newblock \href {http://dx.doi.org/10.1007/JHEP08(2014)063}
  {\path{doi:10.1007/JHEP08(2014)063}}.

\bibitem{JinfengLiao}
J.-f. Liao, {To appear QM2017 proceedings}.

\bibitem{Abelev:2012hxa}
B.~Abelev, et~al., Phys. Lett. B720 (2013) 52--62.

\bibitem{CMS:2012aa}
S.~Chatrchyan, et~al., Eur. Phys. J. C72 (2012) 1945.

\bibitem{Abelev:2012di}
B.~Abelev, et~al., Phys. Lett. B719 (2013) 18--28.

\bibitem{Chatrchyan:2012xq}
S.~Chatrchyan, et~al., Phys. Rev. Lett. 109 (2012) 022301.

\bibitem{Aad2012330}
Physics Letters B 707~(3–4) (2012) 330 -- 348.

\bibitem{Noronha-Hostler:2016eow}
J.~Noronha-Hostler, B.~Betz, J.~Noronha, M.~Gyulassy, Phys. Rev. Lett. 116~(25)
  (2016) 252301.

\bibitem{Betz:2011tu}
B.~Betz, M.~Gyulassy, G.~Torrieri, Phys. Rev. C84 (2011) 024913.

\bibitem{Betz:2012qq}
B.~Betz, M.~Gyulassy, Phys. Rev. C86 (2012) 024903.

\bibitem{Bilandzic:2010jr}
A.~Bilandzic, R.~Snellings, S.~Voloshin, Phys. Rev. C83 (2011) 044913.
\newblock \href {http://arxiv.org/abs/1010.0233} {\path{arXiv:1010.0233}}.

\bibitem{Jia:2012ez}
J.~Jia, Phys. Rev. C87~(6) (2013) 061901.

\bibitem{Betz:2016ayq}
J.~e.~a. Noronha-Hostler, Phys. Rev. C95~(4) (2017) 044901.
\newblock \href {http://dx.doi.org/10.1103/PhysRevC.95.044901}
  {\path{doi:10.1103/PhysRevC.95.044901}}.

\bibitem{Huhn:2017uus}
P.~Huhn, J. Phys. Conf. Ser. 798~(1) (2017) 012075.
\newblock \href {http://dx.doi.org/10.1088/1742-6596/798/1/012075}
  {\path{doi:10.1088/1742-6596/798/1/012075}}.

\bibitem{Chien:2015vja}
Y.-T. Chien, A.~Emerman, Z.-B. Kang, G.~Ovanesyan, I.~Vitev, Phys. Rev. D93~(7)
  (2016) 074030.
\newblock \href {http://dx.doi.org/10.1103/PhysRevD.93.074030}
  {\path{doi:10.1103/PhysRevD.93.074030}}.

\bibitem{Bianchi:2017wpt}
E.~Bianchi, J.~Elledge, A.~Kumar, A.~Majumder, G.-Y. Qin, C.~Shen\href
  {http://arxiv.org/abs/1702.00481} {\path{arXiv:1702.00481}}.

\bibitem{Djordjevic:2016vfo}
M.~Djordjevic, B.~Blagojevic, L.~Zivkovic, Phys. Rev. C94~(4) (2016) 044908.
\newblock \href {http://dx.doi.org/10.1103/PhysRevC.94.044908}
  {\path{doi:10.1103/PhysRevC.94.044908}}.

\bibitem{Andres:2016iys}
C.~Andres, N.~Armesto, M.~Luzum, C.~A. Salgado, P.~Zurita, Eur. Phys. J.
  C76~(9) (2016) 475.

\bibitem{Khachatryan:2016odn}
V.~Khachatryan, et~al.\href {http://arxiv.org/abs/1611.01664}
  {\path{arXiv:1611.01664}}.

\bibitem{Bertens:2017krr}
R.~A. Bertens\href {http://arxiv.org/abs/1704.03028} {\path{arXiv:1704.03028}}.

\bibitem{ATLAS-CONF-2016-105}
{ATLAS Collaboration}, {ATLAS-CONF-2016-105} (2016).
\newblock \href{http://cds.cern.ch/record/2220372}{[link]}.
\newline\urlprefix\url{http://cds.cern.ch/record/2220372}

\bibitem{QWang}
Q.~Wang, {To appear QM2017 proceedings}.

\bibitem{Aad:2015lwa}
G.~Aad, et~al., Phys. Rev. C92~(3) (2015) 034903.

\bibitem{Prado:2016szr}
C.~A.~G. Prado, J.~Noronha-Hostler, A.~A.~P. Suaide, J.~Noronha, M.~G. Munhoz,
  M.~R. Cosentino\href {http://arxiv.org/abs/1611.02965}
  {\path{arXiv:1611.02965}}.

\bibitem{Qian:2017ier}
J.~Qian, U.~Heinz, R.~He, L.~Huo\href {http://arxiv.org/abs/1703.04077}
  {\path{arXiv:1703.04077}}.

\bibitem{Tachibana:2017syd}
Y.~Tachibana, N.-B. Chang, G.-Y. Qin\href {http://arxiv.org/abs/1701.07951}
  {\path{arXiv:1701.07951}}.

\bibitem{Andrade:2014swa}
R.~P.~G. Andrade, J.~Noronha, G.~S. Denicol, Phys. Rev. C90~(2) (2014) 024914.
\newblock \href {http://dx.doi.org/10.1103/PhysRevC.90.024914}
  {\path{doi:10.1103/PhysRevC.90.024914}}.

\bibitem{Nahrgang:2014vza}
M.~Nahrgang, J.~Aichelin, S.~Bass, P.~B. Gossiaux, K.~Werner, Phys. Rev.
  C91~(1) (2015) 014904.

\bibitem{Nahrgang:2016wig}
M.~Nahrgang, J.~Aichelin, P.~B. Gossiaux, K.~Werner, J. Phys. Conf. Ser.
  668~(1) (2016) 012024.

\bibitem{Cao:2013ita}
S.~Cao, G.-Y. Qin, S.~A. Bass, Phys. Rev. C88 (2013) 044907.
\newblock \href {http://dx.doi.org/10.1103/PhysRevC.88.044907}
  {\path{doi:10.1103/PhysRevC.88.044907}}.

\bibitem{Barbano}
A.~Barbano, {To appear QM2017 proceedings}.

\bibitem{YingruXU}
Y.~Xu, {To appear QM2017 proceedings}.

\bibitem{Christiansen:2016uaq}
P.~Christiansen, J. Phys. Conf. Ser. 736~(1) (2016) 012023.

\end{thebibliography}







\end{document}